\documentclass[onecolumn,showpacs]{revtex4}

\usepackage{graphicx}%
\usepackage{dcolumn}
\usepackage{amsmath}

\makeatletter
\def\btt#1{\texttt{\@backslashchar#1}}%
\DeclareRobustCommand\bblash{\btt{\@backslashchar}}%
\makeatother

\begin{document}


\title{USING  INKSURVEY  WITH  PEN-ENABLED  MOBILE  DEVICES FOR  REAL-TIME  FORMATIVE  ASSESSMENT\\*
I. APPLICATIONS  IN  DIVERSE  EDUCATIONAL  ENVIRONMENTS\\*
Technology in Practice Strand
}

\author{F.V. Kowalski}
\affiliation{Physics Department, Colorado
School of Mines, Golden CO. 80401 U.S.A.}

\author{Thomas J. Colling, }
\affiliation{Rancocas Valley Regional High School, Mt. Holly NJ, USA}

\author{J. V. Gutierrez Cuba, }
\affiliation{Universidad de las Am\'{e}ricas Puebla, Mexico}

\author{Enrique Palou, }
\affiliation{Universidad de las Am\'{e}ricas Puebla, Mexico}

\author{Gus Greivel, }
\affiliation{Physics Department, Colorado
School of Mines, Golden CO. 80401 U.S.A.}

\author{Todd Ruskell, }
\affiliation{Physics Department, Colorado
School of Mines, Golden CO. 80401 U.S.A.}

\author{Tracy Q. Gardner, }
\affiliation{Physics Department, Colorado
School of Mines, Golden CO. 80401 U.S.A.}

\author{S.E. Kowalski, }
\affiliation{Physics Department, Colorado
School of Mines, Golden CO. 80401 U.S.A.}

\begin{abstract}

InkSurvey is free, web-based software designed to facilitate the collection of real-time formative assessment.  Using this tool, the instructor can embed formative assessment in the instruction process by posing an open-format question.  Students equipped with pen-enabled mobile devices (tablet PCs, iPads, Android devices including some smartphones) are then actively engaged in their learning as they use digital ink to draw, sketch, or graph their responses.  When the instructor receives these responses instantaneously, it provides insights into student thinking and what the students do and do not know.  Subsequent instruction can then repair and refine student understanding in a very timely manner.

Although this pedagogical tool is appealing because of its broad theoretical foundations, the cost of pen-enabled mobile technology was until recently a significant barrier to widely implementing this teaching model. However, less expensive tablets, iPads, and Android devices are now filling the market (and student backpacks) and greatly lowering that barrier.

To illustrate the wide applicability of this use of technology, we report a series of seven vignettes featuring instructors of diverse subjects (mathematics, food chemistry, physics, biology, and chemical engineering), with students using diverse pen-enabled mobile devices (tablet PCs, iPads, and Android 4.0 tablets), in diverse educational environments (K-12, community college, publicly-funded engineering university, private university, graduate school), in two countries (United States and Mexico).  In a companion paper, each instructor also shares some data, insights, and/or conclusions from their experiences regarding the effectiveness of this tool.

\end{abstract}


\maketitle

\section{PROBLEM  STATEMENT  AND  CONTEXT}
\label{sec:problem}

In the National Research Council's seminal summary of current knowledge about how people learn \cite{bransford1}, there are compelling and repeated calls to embed formative assessment in the learning process.  Based on a broad theoretical foundation, a significant body of research indicates that frequent formative assessment can actively engage learners, effectively inform instruction, and increase student metacognition.  However, educators' attempts to gather formative assessment often prove not only cumbersome, but include painful delays until the instructor is able to respond to misconceptions revealed in the formative assessments.

\section{METHOD  EMPLOYED}
\label{sec:method}

{\em InkSurvey} is web-based software designed to facilitate the collection of real-time formative assessment \cite{kowalski2}; it can be used for free (ticc.mines.edu) and is compatible with pen-enabled mobile technology including tablet PCs, iPads, and Android devices (4.0 and higher).   Using this tool, the instructor can embed formative assessment in the instruction process by posing open-format questions.  By avoiding multiple choice questions, this affords more insightful probing of student understanding. Students then use digital ink to draw, sketch, write, or graph their responses, which actively engages them in their learning.  When the instructor receives these responses instantaneously, it provides insights into student thinking.  Both the students and the instructor have a more accurate realization of what the students do and do not know.  Subsequent instruction can then repair and refine student understanding in a very timely manner and in a climate where students are receptive to these revisions of their understanding.

To illustrate the wide applicability of this use of technology, the following vignettes describe instructors of diverse subjects, with students using diverse pen-enabled mobile devices, in diverse educational environments, in two countries.

\section{RESULTS  AND  EVALUATION:  Seven Vignettes}
\label{sec:results}

\subsection{University Honors Calculus III}
\label{sec:calc}

GG recently introduced {\em InkSurvey} to 87 first-term college students in this course at the Colorado School of Mines, a publicly funded engineering university. This large, lecture-based course is at times very visual, and {\em InkSurvey} affords him the opportunity to engage students in active learning and get student feedback in a much more organic form than is possible with multiple choice questions on clickers. This allows students to make common mistakes, understand that their peers are also making these mistakes (they are common for a reason), and actively engage in a discussion of the misconception that led to these mistakes.  As a consequence, they are able to learn through this real-time formative assessment and repair misconceptions before they become deep-rooted.

The lecture in this vignette involves the introduction of triple integrals over general regions of three-dimensional space as rectangular or box shaped regions in space.  These regions can be defined with inequalities on the coordinate variables (x, y and z) that involve only constant endpoints.  For example a $3x3x2$ box might be defined by the inequalities $0<x\leq 3$, $0\leq y\leq 3$, and $0\leq z\leq 2$. Students are told that in general, regions of integration are bounded by surfaces that may not correspond to constant equations.  After a few examples are presented, students are asked to graph and determine the inequalities necessary to define a tetrahedral region bounded by the coordinate planes and the plane $x+y+2z=2$.\\*
Sample Student {\em InkSurvey} Drawing:

\includegraphics[scale=0.2]{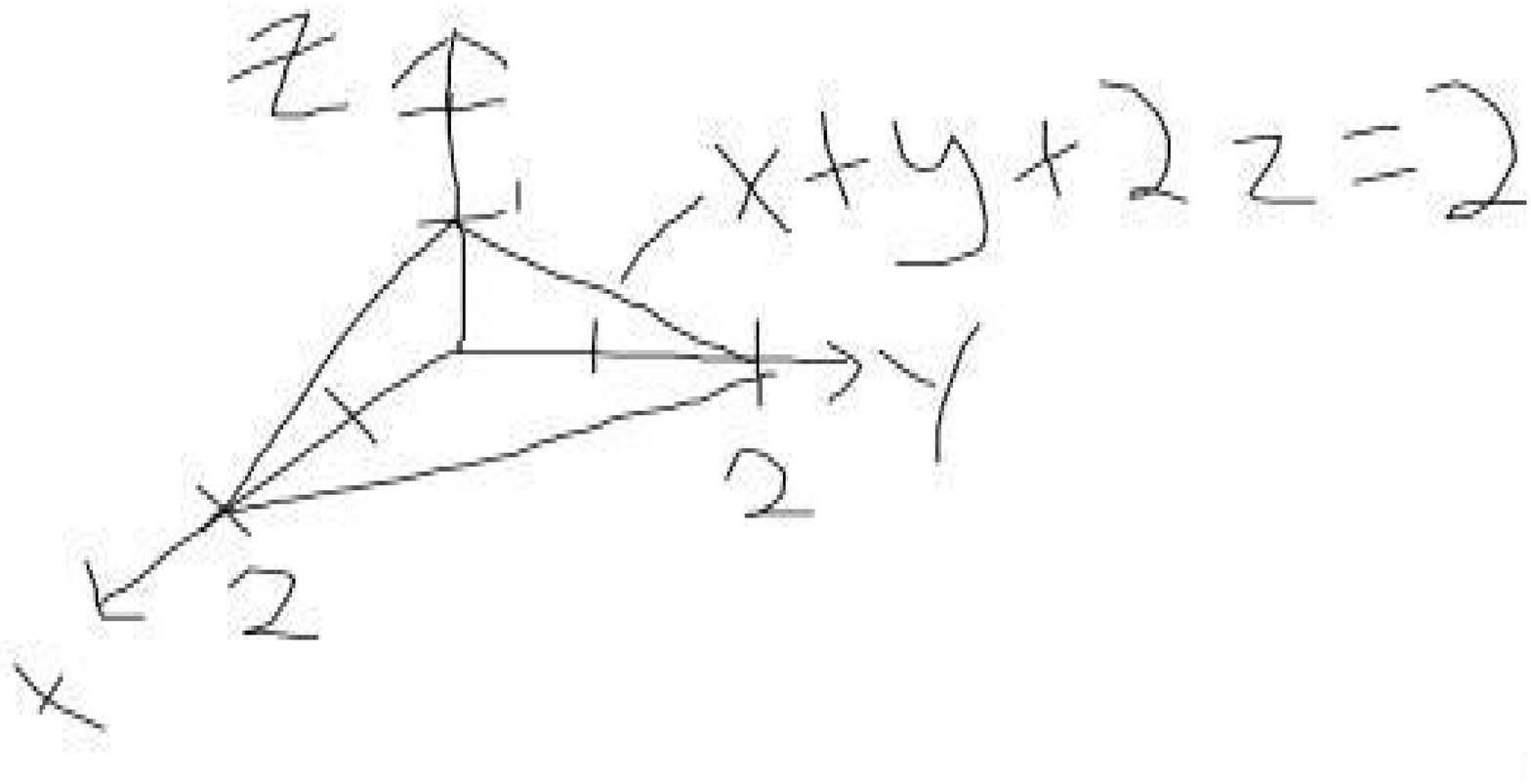}

Students are expected to report the inequalities $0<x\leq 2$ and $0<z\leq 1-x/2-y/2$.  At this point, many students struggle with the common misconception that all of these inequalities will have constant boundary values, based on the original derivation of the triple integral.  This is evidenced by student responses to the following {\em InkSurvey} question:

Q4: Set up the inequalities to define the first octant region bounded by the coordinate planes and the plane
       $x + y + 2z = 2$ as a Type I Region.\\*
Sample {\em InkSurvey} Response (student \#84):

\includegraphics[scale=0.2]{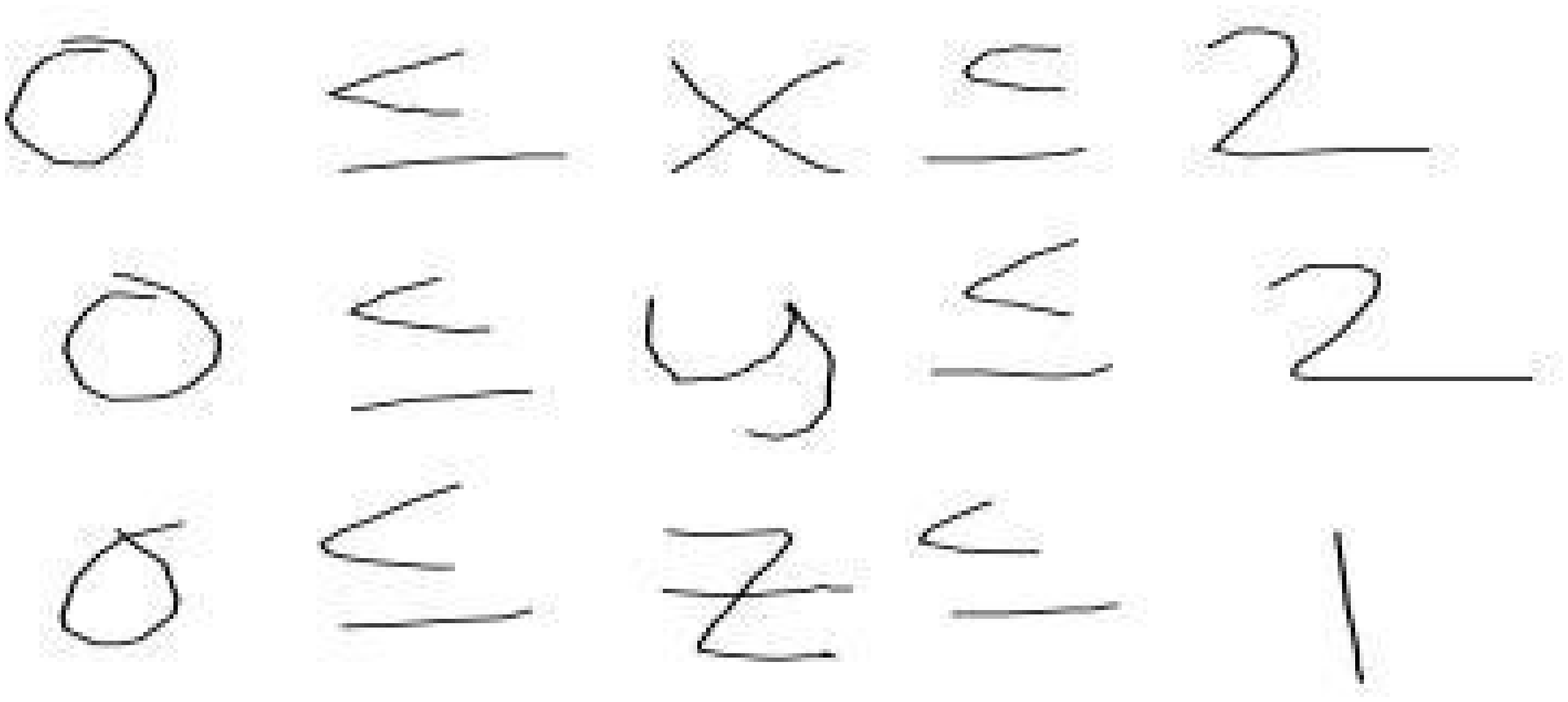}

As in the sample above, 32 out of 83 student responses showed all constant bounds on these inequalities; four of the students didn't respond at all.  Immediately after receiving them, GG then projected some of these student responses and discussed in class an appropriate approach to determining these bounds and defining the correct inequalities.  After this discussion, the students were asked a slight variation of the question on {\em InkSurvey}.  For this question, the students are expected to report the inequalities $0\leq y\leq 2$, $0\leq z\leq 2-y/2$ and $0\leq x\leq 2-y-2z$.

Q6: Set up the inequalities to define the same region as a Type II Region.\\*
Sample {\em InkSurvey} Response (student \#84):

\includegraphics[scale=0.2]{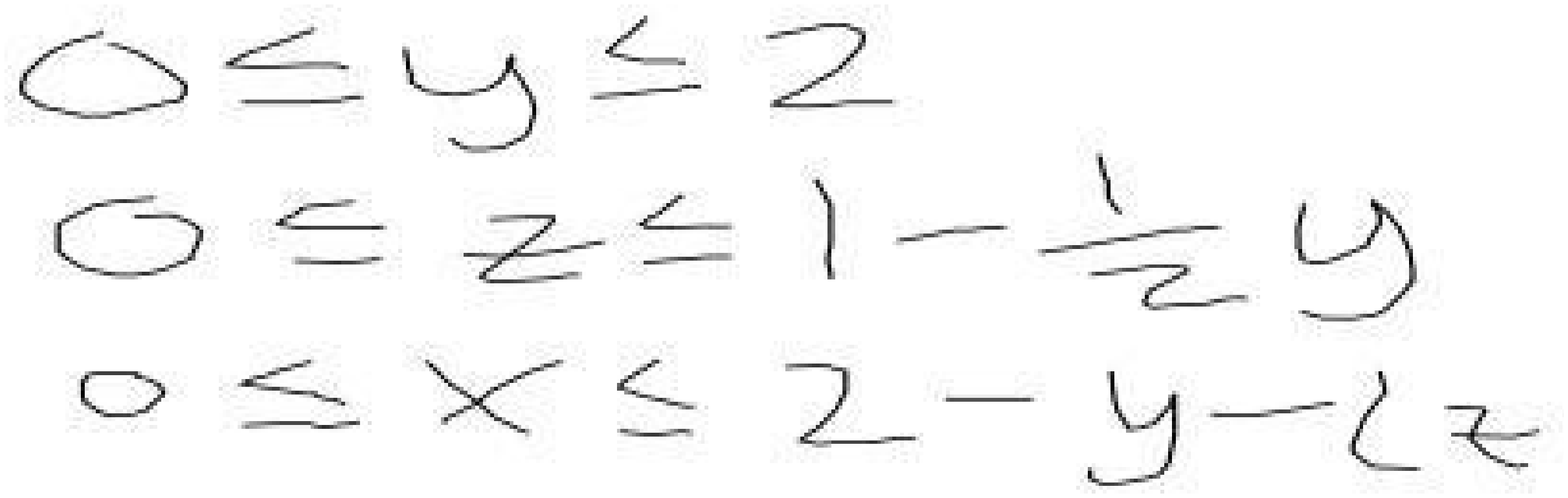}

Taken together, the sample responses above (from student \#84) reveal the real-time progress in a single student's understanding during one class session. When Q6 was posed, all students responded, only 4 still demonstrated the misconception in one dimension, and none of the students provided all constant bounds.

\subsection{High School Algebra}
\label{sec:algebra}

When TC was first exposed to {\em InkSurvey} at a worldwide gathering of educators in New Dehli, he realized it could provide an avenue for using pen-enabled mobile technology to address his public high school's district-wide focus on real-time assessment of standards-based declarative and procedural knowledge of students.  He found it particularly appealing since it closely matched his teaching personality and preferred pedagogical methods.  Now, his high school Algebra I students use {\em InkSurvey} to complete problem-solving exercises and receive both individual and class-wide feedback on a near-daily basis.

Prior to TC's implementation of {\em InkSurvey}, a few selected students in his Algebra classes would show their problem solutions on the board and feedback was given only to those 2 or 3 students. Then, due to time constraints, the class would need to move on to another problem.  With the use of {\em InkSurvey}, all of his students now perform their work on Tablet PCs and submit their solutions. TC quickly reviews all of the answers from the class in order to fix any misconceptions about the lesson prior to students leaving class and practicing the material more on their own at home.  He enjoys receiving the instantaneous feedback from all 25 students, so that he can compare and contrast their solutions and fix as many problems as possible. Also, as TC projects selected solutions that students have submitted, the class can view alternate options in a constructive and beneficial way, and come to realize that there may be multiple correct mathematical methods that are within the properties and definitions of Algebra I.

\subsection{University Upper-level Undergraduate and Graduate Courses in Food Chemistry}
\label{sec:foodchem}

In Mexico, the How People Learn framework \cite{bransford1,bransford2} was applied to redesign two courses in Food Chemistry at Universidad de las Américas Puebla, a prestigious private university. The goal was to improve teaching and learning by creating high-quality learning environments that promote interactive classrooms with formative assessments by means of Tablet PCs and associated technologies \cite{cuba}.  One targeted course is a junior level required course (typically 10-25 students) for food engineering and nutrition BS programs; the other is a first-semester required course (typically 5-10 students) for the food science MSc program and also an elective for the PhD in food science program. In both courses, a major goal is to help students think the way a food chemist does. Instructors of these courses utilize {\em InkSurvey} mainly to pose open-ended formative assessment questions to students during class to gauge student learning in real time, provide immediate feedback, and make real-time pedagogical adjustments as needed.

\subsection{University Introductory Physics (Calculus-Based)}
\label{sec:phys}

TR is using {\em InkSurvey} to instruct a freshman calculus-based Physics I course at Colorado School of Mines.  This course already employed a variety of techniques to engage students, including teaching in a hybrid studio model \cite{furtak} and utilizing personal response systems (clickers) to answer multiple-choice questions during lecture.  Incorporating {\em InkSurvey} with Nexus 7 (Android 4.0) tablets allows TR to explore the next level of real-time formative assessment to more deeply probe student understanding.

In many cases, he has recast traditional clicker questions as open-format questions. For example,
Clicker question: A bowling ball and a ping-pong ball both have the same initial momentum.  If you exert the same constant force on each, how do their stopping distances compare?
A) distance to stop bowling ball $>$ distance to stop ping-pong ball.  B)  distance to stop ping-pong ball $>$ distance to stop bowling ball.  C) It takes the same distance. D) Not enough information to tell.
{\em InkSurvey} version: A bowling ball and a ping-pong ball both have the same initial momentum.  If you exert the same constant force on each, how do their stopping distances compare? Why?

With the open-format version, instead of simply sitting back and "thinking about it" (which is often what students do with the clicker version), students need to justify their answers in writing.  If equations are involved, as here, students are forced to actually work through the problem, solidifying their understanding, or revealing their difficulties, on the tablets during class.  Furthermore, {\em InkSurvey} responses to the above question revealed to TR that some students arrived at the correct conclusion but with faulty reasoning.  Some, for example, displayed thinking in terms of a constant velocity, which is not the case in this situation.  This incorrect path to the correct conclusion may not have been apparent without an open-format question.   After rapidly scanning {\em InkSurvey} responses during class, TR was able to immediately address this misconception.

\section{Community College Introductory Biology Class}
\label{sec:bio}

At Red Rocks Community College, SK works with a broad diversity of learners.  In the recent past, many students were neither comfortable nor fluent in their use of technology to enhance learning, so she was uncertain if {\em InkSurvey} would present special obstacles when she implemented its use.  Using carts of both HP Tablets and iPads and a portable wireless access point in her classroom, the students needed very little "tech support" as they quickly became proficient at submitting responses.

Since many students struggle to "connect the dots" in the context of the class, she has found it particularly useful to use a sequence of questions to scaffold learning and nurture these skills.   For example, the following questions were embedded in a lecture on enzyme activity; they were launched one at a time, with class discussion of the responses interspersed.\\*
1.  Review:  what does pH measure?\\*
2. A lysosome is a special compartment within a eukaryotic cell, filled with destructive enzymes. The internal pH of the lysosome is 4.5. Thinking about the optimal pH of other enzymes we observed in lab this week, draw a graph predicting the optimal pH of one of these enzymes. (pH on the x-axis and the rate of the reaction on the y-axis)\\*
3. Why does this enzyme become less active at a neutral pH? Sketch the shape of the molecule at each pH to illustrate your understanding.\\*
4. Predict what would happen if destructive enzymes of the lysosome leaked into the cytoplasm.

By completing this series of questions, students were engaged in their learning, motivated to participate in the interspersed discussions, and able to effectively relate previously studied topics (lab observations, pH, weak hydrogen bonds, structure and denaturation of proteins, etc.) to the current topic of enzymes and how they work. This provided practice in "connecting the dots" and served as an opportunity for some self-discovery of important relationships and concepts.

\section{University Upper-Level Undergraduate Chemical Engineering and Physics}
\label{sec:chemphys}

TG and FK couple the use of {\em InkSurvey} with interactive computer simulations in their advanced undergraduate chemical engineering and physics courses, respectively. Although there is a wealth of simulations ("sims") available online (many free or associated with textbooks) to help students engage with and better visualize difficult concepts, it is challenging to know how to best use these in STEM classrooms.  Too little guidance from the instructor, and students may not construct correct understandings; too much, and the "cookbook" atmosphere discourages both self-exploration and engagement in the learning process, and may prevent students from understanding concepts at a level necessary to help remember and apply them in other situations.

TG and FK typically have their students explore a sim independently, learning all they can on their own.  Then, the instructors use a series of {\em InkSurvey} questions in class to probe student understanding, identify points of difficulty, and then guide students as they return to the sim for further exploration or testing of concepts and subtleties.  This guided refinement of understanding during the learning process is seamless, since the same devices are used for exploring the simulations, delivering scaffolding questions to guide explorations, and submitting responses to reveal understanding in real time.

\section{University Upper-Level Creativity in Physics}
\label{sec:creativity}

In a course designed to nurture creativity in undergraduate and graduate level physics students, FK successfully utilizes {\em InkSurvey} in group interactions to generate new ideas. First, a problem is stated to the class; then student responses, anonymous to their peers, are requested and displayed in each of the following steps:\\*\\*
(1) Two separate questions, launched simultaneously, solicit individual responses prior to initiation of a class discussion:\\*
a. ideas (unrestrained ideas, not limited to those that students think are practical).\\*
b. questions (about constraints, factual information, motivation for studying the problem, etc.)\\*\\*
As the real-time student responses are collected, the facilitator sorts the ideas and questions into categories. The questions are addressed in a class discussion. The categorized ideas are displayed anonymously to the group and discussed by the group as appropriate. The ideation request can then be repeated if appropriate.\\*\\*
(2) Next, using the organized list of ideas, 2 separate questions, launched simultaneously, solicit:\\*
a. positive critical comments on the ideas, and b. negative critical comments on the ideas.\\*\\*
The submitted critical comments  are organized by the instructor and then presented to the group and discussed as appropriate. This serves as the foundation for the next step.\\*\\*
(3) Students construct and submit a metric to determine the solution based on these positive and negative comments. The final group solution is chosen from these.
This process is then repeated to generate additional new ideas and refine the solution.\\*

Placing all students and ideas on equal footing, {\em InkSurvey} is used to constructively deal with social issues in the process of generating new ideas. It addresses evaluation apprehension since displayed student responses are blind to the audience (but not the instructor). Since their work contains a student identifier, visible only to the instructor, the student is accountable for their participation, mitigating a group performance degradation issue. The anonymity of the responses also encourages participation of each student in the electronic brainstorming process, regardless of gender, race, etc. The request for questions enhances metacognition by requiring students to write about what they do and do not understand. Step 1 requires independent thinking; steps 2 and 3 build on this and force the student to pay careful attention to the ideas of others.

\section{CONCLUSION}
\label{sec:concl}

These seven vignettes demonstrate the wide applicability of this use of technology by instructors of diverse subjects, with students using diverse pen-enabled mobile devices, in diverse educational environments, in two countries.  A companion paper addresses indications of the effectiveness of this pedagogical tool.

\begin{acknowledgments}
Various facets described here have been supported by:  HP Catalyst Program (FK, SK, TC, JC, TG, and EP), NSF grants \#1037519 and \#1044255 (FK, SK, and TG), the Trefny Endowment (GG, TR), and the National Council for Science and Technology of Mexico (CONACyT) (JC).  We appreciate this generous support.
\end{acknowledgments}

\end{document}